\documentclass[runningheads,a4paper]{llncs}

\usepackage{amssymb}
\usepackage{graphicx}
\usepackage{comment}
\usepackage{cite}
\usepackage{cleveref}
\usepackage{algorithmic}
\usepackage[algo2e,ruled,vlined,linesnumbered]{algorithm2e}
\usepackage{booktabs}

\usepackage{url}
\newcommand{\keywords}[1]{\par\addvspace\baselineskip
\noindent\keywordname\enspace\ignorespaces#1}

\usepackage{tikz}

\begin{document}

\mainmatter  

\title{Hybrid Quantum-Classical Neural\\ Architecture Search}

\titlerunning{Hybrid Quantum-Classical Neural Architecture Search}

%
%
\author{Alberto Marchisio\textsuperscript{1,2}, Muhammad Kashif\textsuperscript{1,2}, Nouhaila Innan\textsuperscript{1,2}, and Muhammad Shafique\textsuperscript{1,2}
}

\authorrunning{A. Marchisio, M. Kashif, N. Innan, and M. Shafique}

\institute{
\textsuperscript{1} eBRAIN Lab, Division of Engineering, New York University Abu Dhabi (NYUAD), Abu Dhabi, United Arab Emirates\\
\textsuperscript{2} Center for Quantum and Topological Systems (CQTS), NYUAD Research Institute, NYUAD, Abu Dhabi, United Arab Emirates
}

%
%

\maketitle

\begin{abstract} 
Hybrid quantum-classical neural networks (HQNNs) are emerging as a practical approach for quantum machine learning in the noisy intermediate-scale quantum (NISQ) era, as they combine classical learning components with parameterized quantum circuits in an end-to-end trainable framework. However, their performance and efficiency depend strongly on architectural choices such as data encoding, circuit structure, measurement design, and the coupling between classical and quantum modules. This makes manual design increasingly difficult, especially when hardware limitations and resource constraints must also be taken into account. In this paper, we study the foundations of HQNNs and neural architecture search (NAS), discuss how NAS extends to quantum and hybrid settings, and demonstrate FLOPs-aware search (where FLOPs serve as a proxy for computational complexity), as an important hardware-aware direction for building HQNNs that are not only accurate but also computationally efficient and practically deployable.
\keywords{Hybrid quantum-classical neural networks, quantum machine learning, neural architecture search, hardware-aware optimization, quantum neural networks
}
\end{abstract}

\section{Introduction}

Quantum machine learning (QML) has emerged as a promising research direction at the intersection of quantum computing and artificial intelligence, with the goal of leveraging quantum computational primitives to improve learning systems under certain problem settings~\cite{schuld2015introduction, zaman2023survey,kashif_PP, qi2025quantum,innan2023enhancing,innan2024variational}. Among the different QML formulations, hybrid quantum-classical neural networks (HQNNs) have received particular attention because they combine the flexibility of classical deep learning with the expressive potential of parameterized quantum circuits (PQCs), while remaining compatible with the limitations of near-term quantum devices \cite{kashif2021design}.

In an HQNN, classical components are typically used for data preparation, feature transformation, and prediction, whereas the quantum component acts as a trainable layer embedded within the overall model. This hybrid design is attractive in the noisy intermediate-scale quantum (NISQ) era because it avoids relying on fully fault-tolerant quantum computation and instead enables end-to-end learning with shallow circuits and classical optimization~\cite{preskill2018nisq}. However, the performance of HQNNs depends strongly on architectural choices, including the encoding strategy, the number of qubits, circuit depth, gate composition, entanglement pattern, and the way quantum and classical modules are connected~\cite{zaman2024comparative, ahmed2025noisyhqnns,kashif2023impact, zaman2024studying, ahmed2025comparative, sawaika2025privacy, marchisio2025cutting, ahmed2026gatqnn,vyskubov2026scaling,kashif2023unified}.

These challenges motivate the need for automated design methods tailored to hybrid quantum models. Neural architecture search (NAS), which has been highly successful in classical deep learning, offers a natural foundation for this purpose. When extended to quantum and hybrid settings, NAS can systematically explore candidate architectures and identify models that achieve a better balance between predictive quality and implementation cost than manually crafted designs~\cite{kashif2025faqnas, kashif2026closing, maleki2026qnas}.

In this paper, we analyze HQNNs from the perspective of architecture design and demonstrate that the next stage of progress requires moving from heuristic construction to hardware-aware search. We first review the building blocks of HQNNs and the main concepts behind NAS, then formulate how these ideas extend to quantum settings, and finally present hardware-aware approaches for discovering HQNN architectures that are not only accurate but also efficient and feasible under practical constraints. While we use the term hardware-aware in a broad sense, the present work focuses on computational cost proxies (i.e., FLOPs) that are particularly relevant in simulation-driven HQNN development. Extending this framework to incorporate device-level constraints remains an open challenge.

\section{Background}

\subsection{Hybrid Quantum-Classical Neural Networks}

HQNNs couple conventional neural computation with a PQC inside a single trainable learning pipeline \cite{kashif2022demonstrating}. In this setup, the classical part usually prepares the data and interprets the quantum outputs, while the quantum layer acts as a compact nonlinear transformation whose behavior is controlled by variational gate parameters.

\begin{figure}[htpb]
    \centering
    \includegraphics[width=.8\linewidth]{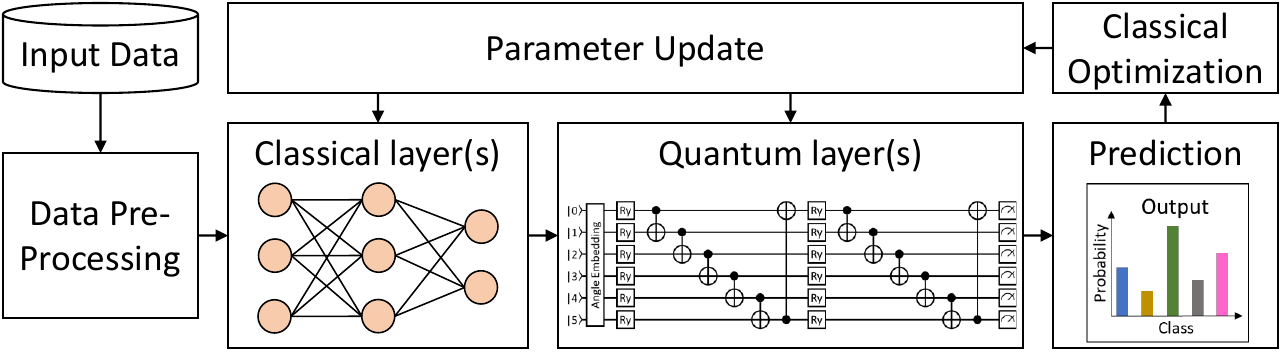}
    \caption{Functionality of a HQNN, composed of both classical layers and quantum layers, equipped with classical optimization and parameter update.}
    \label{fig:HQNN_background}
\end{figure}

Figure~\ref{fig:HQNN_background} illustrates this workflow. An input sample first passes through a data pre-processing stage, followed by one or more classical layers that shape the representation before it is sent to the quantum block. The measured quantum outputs are then returned to the classical domain to form the final prediction, while a classical optimizer updates all trainable parameters across the hybrid model during learning.

Training is performed end-to-end over both sets of parameters, which are composed of both the weights in the classical layers and the variational angles inside the PQC. A standard supervised loss, such as cross-entropy in classification settings, is minimized using classical optimization, while gradients through the quantum block are obtained with differentiable quantum techniques such as the parameter-shift rule. This allows the quantum module to be integrated into familiar deep learning workflows without breaking the overall backpropagation pipeline.

A central practical constraint is that the quantum circuit must remain shallow enough for NISQ-era execution or efficient simulation. For this reason, the number of qubits, the circuit depth, the placement of rotation gates, and the entanglement layout cannot be chosen arbitrarily. Circuits that are too large may introduce unnecessary cost and unstable optimization due to barren plateaus, whereas overly simple circuits may fail to capture useful structure in the data \cite{kashif2024alleviating,McClean_2018arxiv_BarrenPlateaus,kashif2024resqnets}.

This sensitivity to architectural choices is one of the main reasons HQNN design remains challenging. Even small modifications in encoding strategy, gate arrangement, or readout design can noticeably affect performance, efficiency, and robustness~\cite{zaman2024studying, el2025robqfl, el2025designing, ahmed2025comparative, el2024advqunn}. \textit{As a result, systematic methods for exploring hybrid architectures under resource constraints are becoming increasingly important, especially when the characteristics of the target backend can alter the effective behavior of the quantum layer.}

\subsection{NAS Techniques}

\begin{figure}[htpb]
    \centering
    \includegraphics[width=0.8\linewidth]{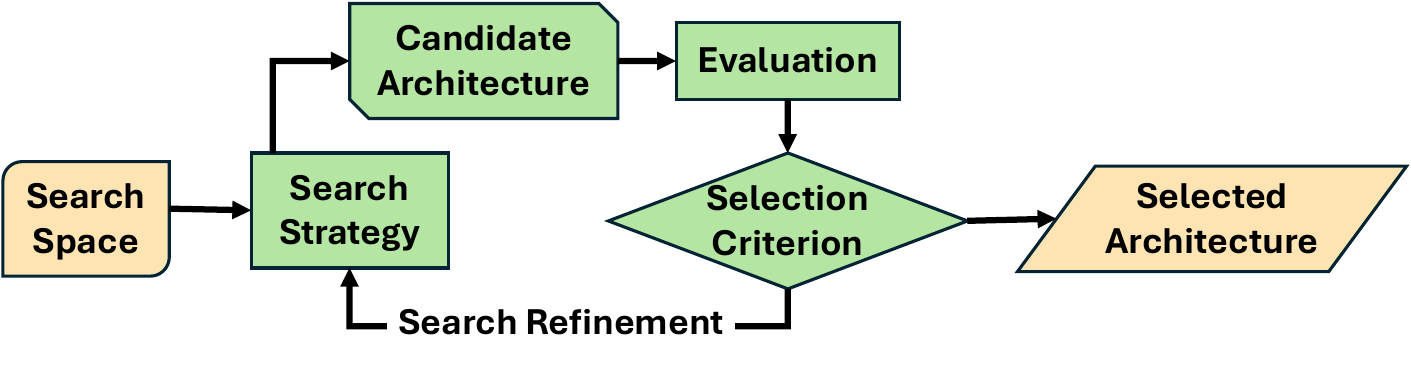}
    \caption{Overview of the standard NAS workflow, including the search space, search strategy, candidate evaluation, and architecture selection.}
    \label{NAS-B}
\end{figure}

NAS automates the design of neural networks by exploring candidate architectures within a predefined search space and selecting those that best satisfy a target objective \cite{elsken2019neural}. As shown in Fig. \ref{NAS-B}, a standard NAS framework is typically composed of three elements: the search space, which defines the candidate models; the search strategy, which guides the exploration process; and the evaluation procedure, which assesses candidate architectures according to performance and efficiency criteria. In classical deep learning, NAS has become a useful approach for reducing manual design effort and enabling a more systematic development of high-performing models. This general NAS framework serves as the starting point for Quantum NAS, where architecture search is adapted to the unique characteristics of quantum and hybrid models.
\subsection{Evaluation Metrics}
We consider both predictive performance and computational efficiency as key evaluation criteria. In standard NAS, the primary objective is typically accuracy, which reflects how reliably a model performs on a given task. Accuracy measures the proportion of correct predictions and serves as a direct indicator of the model’s generalization capability on unseen data. Optimizing for accuracy encourages the search process to favor architectures that maximize task performance.
However, in hardware-aware NAS, relying solely on accuracy can lead to architectures that are costly to deploy on actual hardware~\cite{benmeziane2021comprehensive, marchisio2020nascaps}. To address this, we use Floating Point Operations (FLOPs) as an additional objective to explicitly capture computational complexity, which quantifies the number of arithmetic operations required, providing a hardware-agnostic proxy for execution cost, latency, and energy consumption~\cite{kashif2024computational,Guo:2021,Chen:2023,hsia:2021}. We define quantum FLOPs as the number of floating-point operations required to simulate PQCs on classical hardware, including state-vector updates induced by gate operations. While this metric reflects simulation cost, it differs from hardware-level metrics such as gate count or circuit depth.
By incorporating FLOPs into the optimization process, the search is guided toward architectures that achieve a favorable trade-off between performance and hardware efficiency. This enables the identification of models that are not only accurate but also computationally practical, which is particularly important in resource-constrained or real-world deployment scenarios.


\section{Quantum Neural Architecture Search}
Building on the foundations of HQNNs and NAS, Quantum NAS emerges as an important direction for automating the design of quantum learning systems \cite{du2022quantum, chen2025quantum, agha2025neural}. Quantum NAS provides a systematic way to explore quantum and hybrid architectures under task-specific objectives and resource constraints. Choices such as the number of qubits, circuit depth, gate composition, entanglement structure, and the coupling between quantum and classical modules can significantly affect both predictive accuracy and computational efficiency \cite{innan2024financial,el2026comparative,choudhary2025hqnn,innan2025qnn}. As a result, the design of quantum learning architectures is still largely manual, relying on expert intuition, repeated experimentation, and task-specific heuristics \cite{innan2025circuithunt}. 

Quantum NAS addresses this limitation by shifting architecture construction from handcrafted design to systematic optimization. Instead of fixing a predefined ansatz, Quantum NAS explores candidate quantum architectures and aims to identify those that are best suited to the target task under relevant constraints. In this sense, Quantum NAS is both a design automation tool and a framework for understanding how quantum architectural choices influence model behavior, efficiency, and scalability. 

\subsection{Problem Formulation}
Quantum NAS can be formulated as the problem of selecting an architecture $a \in \mathcal{A}$ from a search space $\mathcal{A}$ such that the resulting model optimizes a task-specific objective:
\begin{equation}
    a^* = \arg\min_{a \in \mathcal{A}} \mathcal{J}(a).
\end{equation}
where $\mathcal{J}(a)$ may correspond either to a single performance criterion or to a combination of objectives. In this work, we treat $\mathcal{J}(a)$ as a multi-objective criterion that reflect both learning quality and implementation cost. In HQNNs, this often involves balancing predictive performance with complexity-related considerations, such as simulation runtime or resource usage.

Unlike classical NAS, Quantum NAS must additionally account for quantum-specific structure. This makes the search problem broader and often more constrained. Evaluating a candidate architecture may require repeated training and simulation of quantum circuits, which can quickly become computationally expensive. Therefore, Quantum NAS is usually framed as a constrained and often multi-objective optimization problem.

\subsection{Search Space in Quantum Models}
The search space in Quantum NAS defines the set of candidate quantum or hybrid architectures that the search algorithm is allowed to explore. Compared with many classical NAS settings, this space is often broader because it includes not only conventional architectural choices, but also quantum-specific design variables. Typical choices include the number of qubits, the number of variational layers, the gate set, the entanglement pattern, the data encoding strategy, the measurement configuration, and the observable used for readout. In hybrid quantum-classical models, the search space may also include the positions, sizes, and types of classical layers, as well as the interface between classical preprocessing and quantum computation.

The search space may be decomposed as 
\begin{equation}
\mathcal{A} =
\mathcal{A}_{\mathrm{enc}} \times
\mathcal{A}_{\mathrm{var}} \times
\mathcal{A}_{\mathrm{meas}} \times
\mathcal{A}_{\mathrm{post}},
\end{equation}
where $\mathcal{A}_{\mathrm{enc}}$ denotes the set of data-encoding choices, $\mathcal{A}_{\mathrm{var}}$ the trainable variational circuit structure, $\mathcal{A}_{\mathrm{meas}}$ the measurement or readout design, and $\mathcal{A}_{\mathrm{post}}$ the classical post-processing or hybrid prediction head. Not all Quantum NAS methods optimize all of these components jointly, but this decomposition highlights that architecture search in quantum machine learning extends beyond selecting a sequence of quantum gates.

Within the quantum part of the model, the search space can be defined at different structural levels. In a gate-wise search space, individual one-qubit and two-qubit operations are selected position by position, providing high flexibility but also leading to a highly combinatorial search problem. In contrast, layer-wise or block-wise search spaces operate on predefined building blocks, such as hardware-efficient layers, entangling templates, or repeated circuit layers, which reduce the search complexity and introduce useful structural bias. Other formulations further consider qubit-wise assignment strategies or macro/micro search spaces, where reusable subcircuits are first identified at a smaller scale and then transferred to larger architectures \cite{martyniuk2024quantum,chen2024qubit,wu2023quantumdarts}. 

\subsection{Search Strategies for Quantum NAS}
Since exhaustive evaluation quickly becomes impractical even for moderately sized quantum circuits, Quantum NAS relies on search strategies that efficiently explore the architecture space. Existing approaches can be grouped into several categories, each offering a different balance between flexibility, computational cost, and search efficiency. 

Evolutionary and genetic methods represent one of the most natural approaches for Quantum NAS because quantum circuit design is discrete \cite{ding2022evolutionary,zhang2023evolutionary,kashif2025faqnas}. In these methods, each candidate architecture is encoded as a circuit genotype, such as a gate sequence, a block configuration, or a graph representation, and a population of such candidates is iteratively improved through mutation, crossover, and selection. Their main advantage is that they can directly operate on combinatorial search spaces and naturally support multi-objective optimization, for example, when balancing predictive performance with circuit depth, gate count, or noise sensitivity. This makes them particularly suitable when the search space includes structural constraints that are difficult to relax into continuous variables.

Reinforcement learning-based methods form a second major category and are especially appealing when circuit design is considered as a sequential construction problem \cite{kuo2021quantum,dutta2025qas}. In this setting, an agent interacts with an environment that represents the partially built quantum circuit and chooses actions such as adding a gate, selecting an entangling pattern, or inserting a circuit block. The policy is then optimized using rewards derived from task performance, resource efficiency, or other search objectives. This formulation is well matched to Quantum NAS because the effect of an architectural decision often depends on the sequence of previous design choices. Related methods, such as Monte Carlo tree search, also exploit this sequential structure by expanding candidate architectures step by step while balancing exploration and exploitation.

A third category consists of differentiable methods, which aim to make the search process compatible with gradient-based optimization \cite{zhang2022differentiable,HE2024106508,11250362, chen2025differentiable}. Instead of selecting one discrete operator at each architectural position, these approaches assign continuous weights or soft probabilities over a set of candidate gates, layers, or blocks. The architecture parameters are then optimized jointly with the circuit parameters during training, and the final relaxed representation is projected back to a discrete architecture after the search stage. Differentiable Quantum NAS is often more sample-efficient than purely discrete search because it avoids evaluating each candidate independently. However, the continuous relaxation may change the original combinatorial problem and can bias the final architecture toward the chosen parameterization. 

A fourth category focuses on reducing the cost of architecture evaluation rather than changing the outer search mechanism itself. One-shot and supernet-based methods train an over-parameterized supercircuit once and allow candidate subcircuits to inherit weights, which avoids retraining every architecture from scratch. Predictor-guided approaches instead learn surrogate models that estimate the quality of candidate circuits directly from their structure, using neural predictors, graph-based encoders, or training-free proxies such as expressibility or noise-related indicators \cite{zhang2021neural}. Bayesian optimization can also be placed in this category when used with a surrogate model and an acquisition rule to identify informative candidates under limited evaluation budgets \cite{choudhary2025graph}. More recent latent-space methods further reduce search complexity by embedding circuits into a compact representation space and performing optimization there rather than directly in the original architecture space. 

Many Quantum NAS frameworks combine elements from multiple categories. For example, reinforcement learning or Bayesian optimization may be used together with predictor-based ranking, while evolutionary search may be constrained by a structured search space or supported by fast proxy evaluation. This hybridization is particularly important in Quantum NAS, where the search space is discrete, the evaluation cost is high, and architecture quality must often be assessed under multiple competing criteria.
\section{Toward Hardware-Aware Quantum NAS: A FLOPs-Aware Perspective}
The design of HQNNs is rapidly emerging as a critical challenge in QML, and existing HQNN design approaches remain largely manual, heuristic, and disconnected from the limitations of NISQ quantum hardware. 
We anticipate that the next stage of progress requires a paradigm shift toward hardware-aware neural architecture search (HANAS), where HQNNs are automatically discovered rather than heuristic-based manual design. 
In this setting, HQNN architecture design must move beyond accuracy-centric optimization and should be focused on a multi-objective perspective that jointly considers accuracy and computational constraints. 
Unlike classical NAS, HQNN search operates over a fundamentally heterogeneous space, where classical layers interact with PQCs, both of which have fundamentally different target hardware. In this section, we focus on FLOPs-aware NAS as a practical instantiation of hardware-aware HQNN design, particularly relevant for simulation-based workflows in the NISQ era. We present a concrete approach where HANAS for HQNNs is performed to identify architectures which are both accurate and are computationally feasible.  

\subsubsection{FLOPs-aware NAS for HQNNs}

The study in~\cite{kashif2025faqnas} aims to address an important question from HANAS for HQNNs perspective, which states that ``\textit{how
can we design architectures that consume minimum computational resources while still achieving competitive performance''?}. The idea is built on the argument that due to the limitations of current NISQ devices, most HQNNs, in practice, are trained on classical simulators, where computational cost can be measured in FLOPs (as a proxy for HANAS for HQNNs), which directly dictates scalability, memory usage, and runtime. Additionally, as hybrid systems increasingly depend on classical infrastructure, the classical resource overhead continues to remain a bottleneck, particularly in data processing and iterative optimization. Therefore, FLOPs-aware design principles have important implications for future hybrid systems. 
In this context, FLOPs provide a practical and hardware-agnostic measure of efficiency, enabling the design of HQNNs that are both scalable and deployable. 
Building on this, the HQNN architecture search is formulated as a multi-objective optimization problem that simultaneously minimizes computational cost and maximizes accuracy. Genetic algorithm, which incorporates computational cost in terms of FLOPs and model accuracy as objectives into the architecture search process, which then enables the identification of models that are not only accurate but also resource-efficient and scalable.
The approach is validated across different benchmark datasets of varying complexity, spanning small, medium, and large-scale tasks, demonstrating its ability to identify efficient and high-performing architectures across diverse problem settings. 

\begin{figure}[t!]
    \centering
    \includegraphics[width=1.0\linewidth]{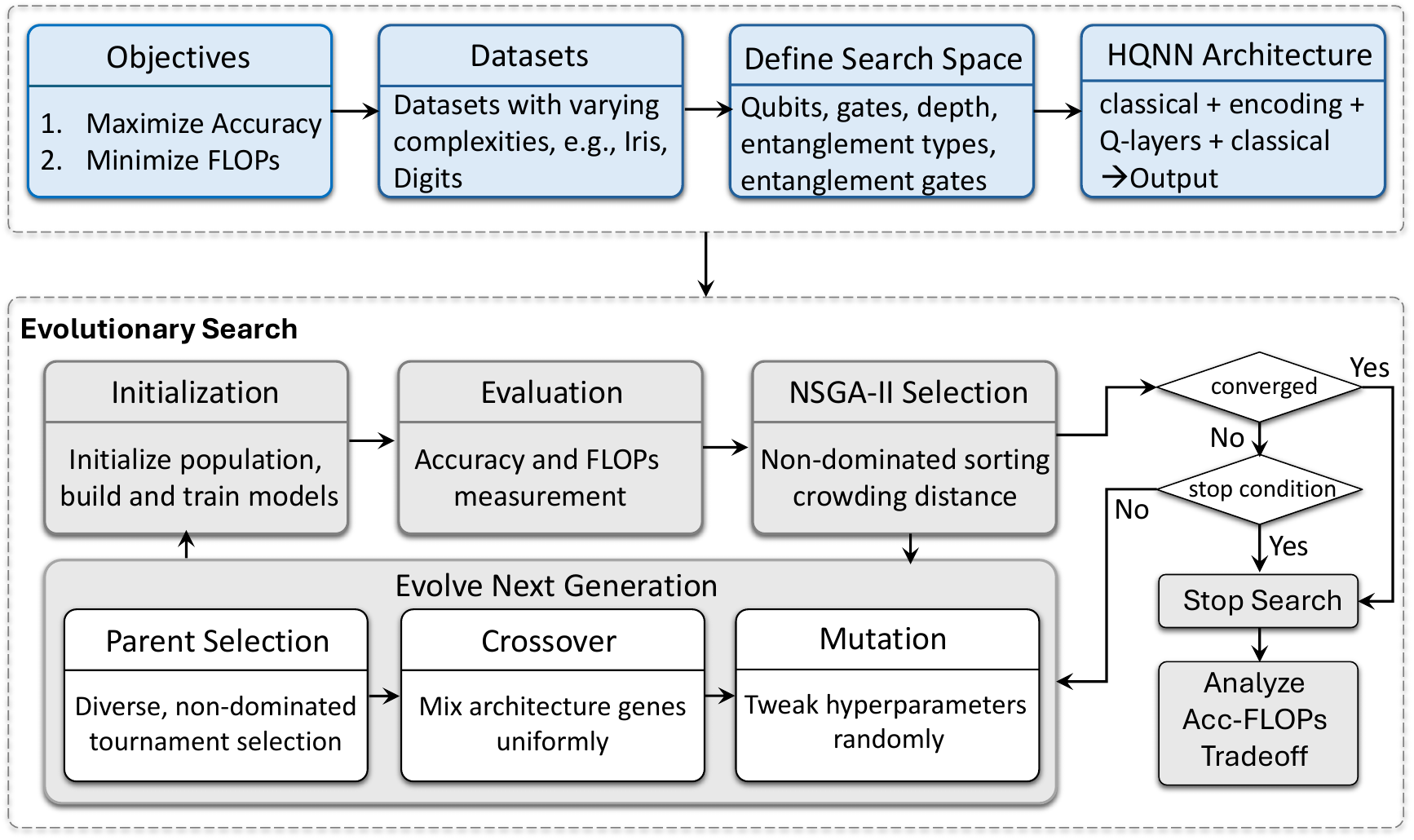}
    \caption{Detailed methodology for FLOPs-aware HQNN architecture search.}
    \label{fig:method_faqnas}
\end{figure}

\paragraph{Proposed Approach and Methodology} The proposed approach utilizes the genetic algorithm (NSGA-II) to explore a structured search space of quantum circuit configurations. The search space includes the number of qubits (\{2,3,..10\}), data encoding methods (\{angle, amplitude\}), parameterized rotation gates (\{$R_x$,$R_y$,$R_z$\}), entangling gate types (\{CNOT, CZ\}), entanglement topology (\{linear, circular\}), and circuit depth (\{1,2,3,4\}).  The classical pre and post processing layers are kept fixed to better analyze the impact of quantum layers towards accuracy and hardware resource requirement. Based on the search space definition above, a total of 23,328 hybrid architectures are evaluated for each dataset.
Each candidate architecture is instantiated as a hybrid model, combining classical pre-processing, a parameterized quantum circuit (built from the search space defined above), and classical post-processing layers. During evaluation, the computational contribution of the quantum component is isolated by measuring quantum FLOPs separately from classical FLOPs, enabling a fair and hardware-relevant comparison across architectures. The search process iteratively evolves candidate solutions using crossover ($p=0.8$), mutation ($p=0.2$), and Pareto-optimal architectures that balance accuracy and computational cost. The search stops evaluating further architectures if there is no improvement in the objectives for two consecutive generations. 
An overview of the complete methodology for FLOPs-aware NAS for HQNNs is presented in Fig.~\ref{fig:method_faqnas}.

\paragraph{Results and Discussion}

\begin{figure}[h]
    \centering
    \includegraphics[width=1.0\linewidth]{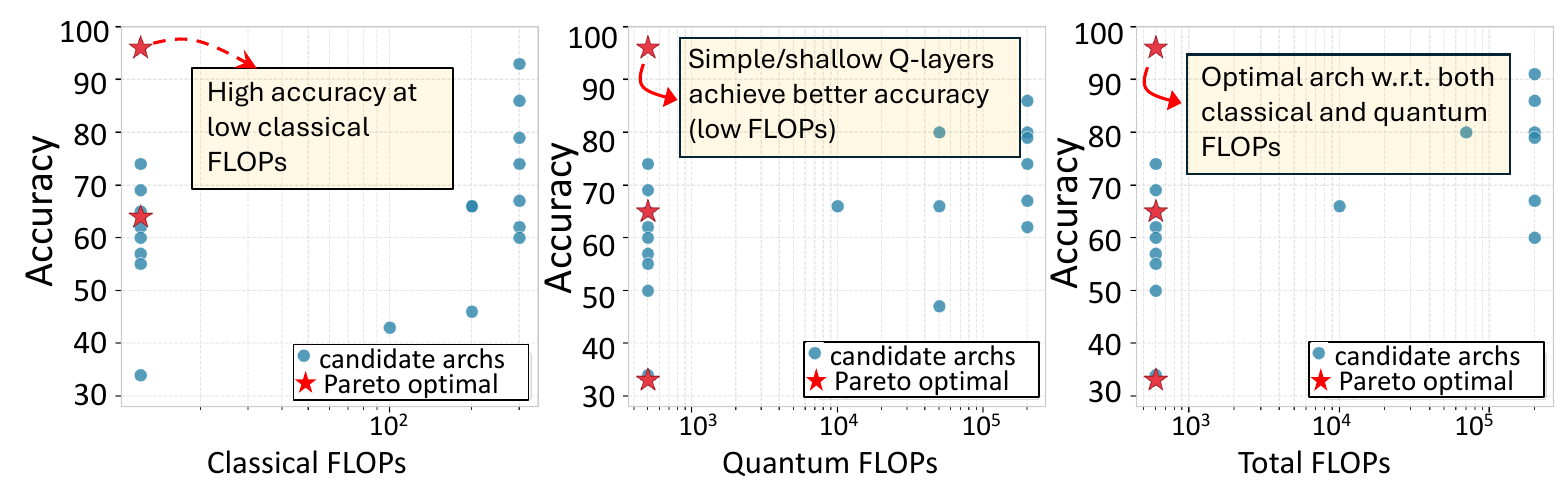}
    \caption{Iris Dataset: Accuracy versus FLOPs for all candidate HQNN architectures (blue dots), and pareto-optimal architectures (red stars). Classical FLOPs vs accuracy (left), quantum FLOPs vs accuracy (middle), and Total FLOPs vs accuracy (right).}
    \label{fig:res_faqnas_iris}
\end{figure}

The experimental evaluation across multiple benchmark datasets reveals several important insights. We present the results for two datasets, namely iris and Digits datasets~\cite{pedregosa2011scikit}. 
For the Iris dataset (Fig.~\ref{fig:res_faqnas_iris}), accuracy varies significantly across candidate architectures, ranging from approximately 40\% to above 90\%.

It can be observed that several Pareto-optimal architectures achieve high accuracy at very low classical FLOPs, indicating that classical overhead is not a limiting factor. The quantum FLOPs analysis shows that strong performance can be attained at both low and high FLOPs budgets, although mid-range configurations exhibit inconsistent behavior. The total FLOPs distribution highlights the importance of careful architecture selection, as FLOPs-aware optimization identifies efficient models that avoid unnecessary computational cost while maintaining high accuracy. Overall, these results suggest that, for simple datasets like Iris, performance is largely governed by the choice of quantum circuit.

\begin{figure}
    \centering
    \includegraphics[width=1.0\linewidth]{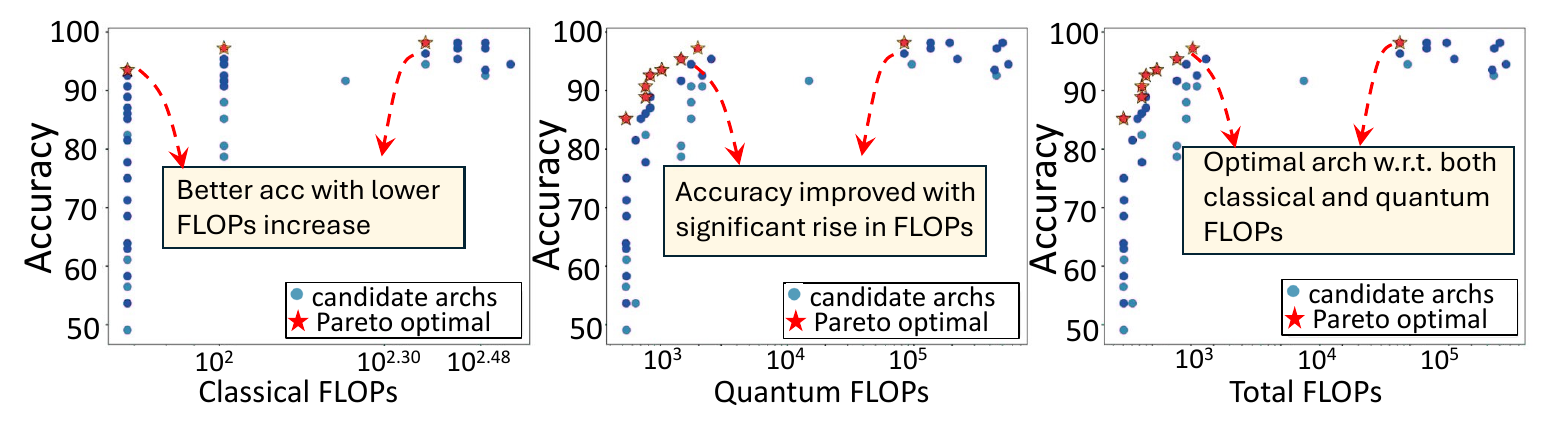}
    \caption{Digits Dataset: Accuracy versus FLOPs for all candidate HQNN architectures (blue dots), and pareto-optimal architectures (red stars). Classical FLOPs vs accuracy (left), quantum FLOPs vs accuracy (middle), and Total FLOPs vs accuracy (right).}
    \label{fig:res_faqnas_digits}
\end{figure}

The Digits dataset, on the other hand, exhibits a slightly different trend (Fig.~\ref{fig:res_faqnas_digits}), accuracy improves rapidly with a small increase in quantum FLOPs, exceeding 90\% at budgets of approximately $10^3$–$10^4$, and quickly saturating near 100\%, beyond which larger quantum circuits offer no benefit. Classical FLOPs show weak correlation with accuracy, with Pareto-optimal solutions appearing across both low and high budgets. 
In contrast, the total FLOPs distribution reveals a well-defined Pareto front, indicating that many architectures achieve optimal performance at significantly lower computational cost. Overall, these results demonstrate that for moderately complex datasets like Digits, FLOPs-aware search effectively identifies compact, high-performing HQNN architectures without unnecessary resource usage.

Overall, the accuracy improvements in HQNNs are primarily driven by quantum FLOPs, while classical FLOPs remain largely fixed due to the constrained design of classical layers. Second, the results consistently exhibit clear Pareto-optimal trade-offs between accuracy and computational cost, demonstrating that high performance can often be achieved without needing the most resource-intensive architectures. Third, there are dataset-dependent thresholds beyond which increasing FLOPs yields diminishing returns, highlighting the risk of overly complex quantum circuits. These findings establish FLOPs-awareness as a practical design principle for HQNNs, enabling more efficient and scalable model selection in the NISQ regime.
At the same time, although FLOPs-aware NAS provides an important step toward systematic HQNN design, it also exposes deeper challenges that cannot be addressed by FLOPs alone, and quantum-hardware specific constraints such noise and transpilation~\cite{kashif2026late} can effect the performance of underlying PQCs in HQNN design.
To this end, how quantum-hardware specific constraints impact the hardware feasibility of HQNNs will typically be interesting to analyze. We confidently envision that the next generation of HANAS frameworks for HQNNs should extend beyond FLOPs to incorporate these dimensions into a unified, multi-objective optimization paradigm. Such frameworks would enable the automated co-design of quantum circuits and classical components under realistic hardware constraints, bridging the gap between abstract model design and deployable quantum systems.


\section{Conclusion}

Hybrid quantum-classical neural networks offer a promising path for near-term quantum machine learning, but their effectiveness depends heavily on architectural decisions that are still often made manually. As HQNNs become more complex, this design process becomes increasingly difficult to scale and less suited to realistic hardware constraints.

In this paper, we positioned Quantum NAS as a systematic approach for exploring HQNN design choices and demonstrated that hardware-aware search is essential for the next stage of progress. By moving beyond accuracy-only optimization and incorporating efficiency and implementation constraints, hardware-aware HQNN search can help bridge the gap between conceptual hybrid models and quantum learning systems that are practical to train and deploy. We considered FLOPs as a proxy to estimate computational resources needed. Although FLOPs can efficiently estimate the computational needs, but they are most feasible for simulator based design. In future, we aim to incorporate the constraints of actual quantum hardware such noise, decoherence, and qubit connectivity etc, within the HANAS framework.

\section*{Acknowledgments.} 

This work was supported in part by the NYUAD Center for Quantum and Topological Systems (CQTS), funded by Tamkeen under the NYUAD Research Institute grant CG008.


\bibliographystyle{ieeetr}
\bibliography{main.bib}

\end{document}